\documentclass{article}

\usepackage{template-iclr2027/iclr2027_conference,times}
\usepackage{hyperref}
\usepackage{url}
\usepackage{graphicx}
\usepackage{caption}
\usepackage{enumitem}
\usepackage{algorithm}
\usepackage{algorithmic}
\usepackage{booktabs}
\usepackage{makecell}
\usepackage{multirow}
\usepackage{amsmath}
\usepackage{amssymb}
\usepackage{fvextra}
\usepackage{float}
\usepackage{placeins}

\DeclareUnicodeCharacter{2265}{\ensuremath{\geq}}
\DeclareUnicodeCharacter{2264}{\ensuremath{\leq}}
\DeclareUnicodeCharacter{2212}{\ensuremath{-}}
\DeclareUnicodeCharacter{2013}{--}
\DeclareUnicodeCharacter{2014}{---}
\DeclareUnicodeCharacter{00D7}{\ensuremath{\times}}
\DeclareUnicodeCharacter{2248}{\ensuremath{\approx}}
\DeclareUnicodeCharacter{26A0}{[WARNING]}
\DeclareUnicodeCharacter{FE0F}{}

\newcommand{\runinhead}[1]{\par\noindent\textbf{#1}\ }
\newcommand{\papertablestyle}{\small\renewcommand{\arraystretch}{1.00}\setlength{\tabcolsep}{4pt}}

\title{RedEvoAgent: Automatic Red-Teaming Agent\\
with Experience-Driven Skill Evolution}

\author{
\hspace*{-\tabcolsep}%
\makebox[0pt][l]{\makebox[\textwidth][c]{%
\begin{tabular}[t]{@{}c@{}}
\textbf{Junjie~Zhang}$^{1}$ \quad
\textbf{Hui~Liu}$^{1*}$ \quad
\textbf{Kecheng~Chen}$^{1}$ \\
\textbf{Xianbo~Mo}$^{2}$ \quad
\textbf{Changsheng~Chen}$^{2}$ \quad
\textbf{Haoliang~Li}$^{1}$ \\[0.6em]
\mdseries
\textsuperscript{1}\,City University of Hong Kong \quad
\textsuperscript{2}\,Shenzhen MSU-BIT University
\end{tabular}}}%
}

\iclrfinalcopy
\begin{document}
\maketitle
\lhead{Preprint}
\renewcommand{\thefootnote}{\fnsymbol{footnote}}
\footnotetext[1]{Corresponding author.}
\renewcommand{\thefootnote}{\arabic{footnote}}

\begin{abstract}
LLM-based agents are increasingly deployed in product-level execution harnesses, where jailbreaks can trigger harmful tool use and persistent state changes, creating greater risks than unsafe text generation alone.
Existing automatic red-teaming methods often rely on fixed attacks, while recent agentic attackers coordinate multiple jailbreak tools and show stronger potential through trajectory-based retrieval.
However, such retrieval can reuse misleading experiences due to retrieval bias and unclear tool credit, and full trajectories add context overhead while reducing interpretability.
We propose RedEvoAgent, a black-box red-teaming agent that distills cross-case attack trajectories into a concise, human-readable attack skill.
The attack skill adaptively evolves through tool-effectiveness profiling and Deciding-Tool Attribution for skill updates, and a validation ratchet that retains only updates improving validation performance.
Experiments on multiple benchmarks, target models, and target execution harnesses show that RedEvoAgent outperforms fixed and agentic baselines, improves tool efficiency, and transfers across attacker models and target execution harnesses.

\end{abstract}

\section{Introduction}
\label{sec:intro}


LLM-based agents are increasingly integrated into product-level harnesses such as Claude Code~\citep{anthropic2026claudecode} and Codex~\citep{openai2025codex}, which allow them to modify local files, transfer data, and call external APIs~\citep{yao2023react,yang2024sweagent,wang2024llmagents,yehudai2026survey}.
These capabilities broaden the security implications of jailbreaks, as successful attacks may cause destructive tool use and persistent changes to system state rather than unsafe text generation~\citep{greshake2023notwhat,ruan2024toolemu,andriushchenko2024agentharm}.
Therefore, product-level black-box agents require adversarial evaluation to characterize their security robustness.
Automatic red-teaming provides a scalable approach for continuous evaluation by simulating real-world malicious attacks, adversarial inputs, and exploit attempts~\citep{perez2022redteaming,ganguli2022redteaming,mazeika2024harmbench}.

Most existing jailbreak methods rely on a fixed attack mechanism.
Some transform or rewrite the original attack prompt to circumvent an agent’s safety-alignment defenses, such as FlipAttack~\citep{liu2024flipattack}, whereas others optimize adversarial prompts within a predefined search space, as in GCG~\citep{zou2023gcg} and AutoDAN~\citep{liu2024autodan}.
Because such methods explore only a limited subset of the attack space, they may miss vulnerabilities exposed by alternative mechanisms and are more likely to fail.
Recent work, therefore, investigates agentic automatic red-teaming, in which an LLM-driven attacker agent leverages prior knowledge about existing attack methods to adaptively coordinate them within a unified workflow.
For example, RedCodeAgent~\citep{guo2026redcodeagent} models attack methods as sequentially invocable tools and utilizes semantic similarity to retrieve successful trajectories as priors to guide the attacker agent’s subsequent decisions.

Despite its effectiveness, trajectory-based retrieval has several limitations. 
First, this paradigm assumes that all retrieved trajectories are useful, which may not hold because semantic retrieval can introduce bias, and the contribution of any individual tool-choice decision to final attack success is often unclear. 
Consequently, irrelevant or misleading experiences may be reused, causing unstable optimization and degraded attack performance. Second, full-trajectory contexts introduce substantial overhead by consuming context-window budget and increasing inference cost. 
Their low-level action records are also difficult for humans to interpret and monitor, hindering auditing over the evolution of strategy.

\begin{figure}[H]
    \centering
    \captionsetup{aboveskip=4pt, belowskip=8pt, skip=2pt}
    \vspace{-2mm}
    \includegraphics[width=\textwidth]{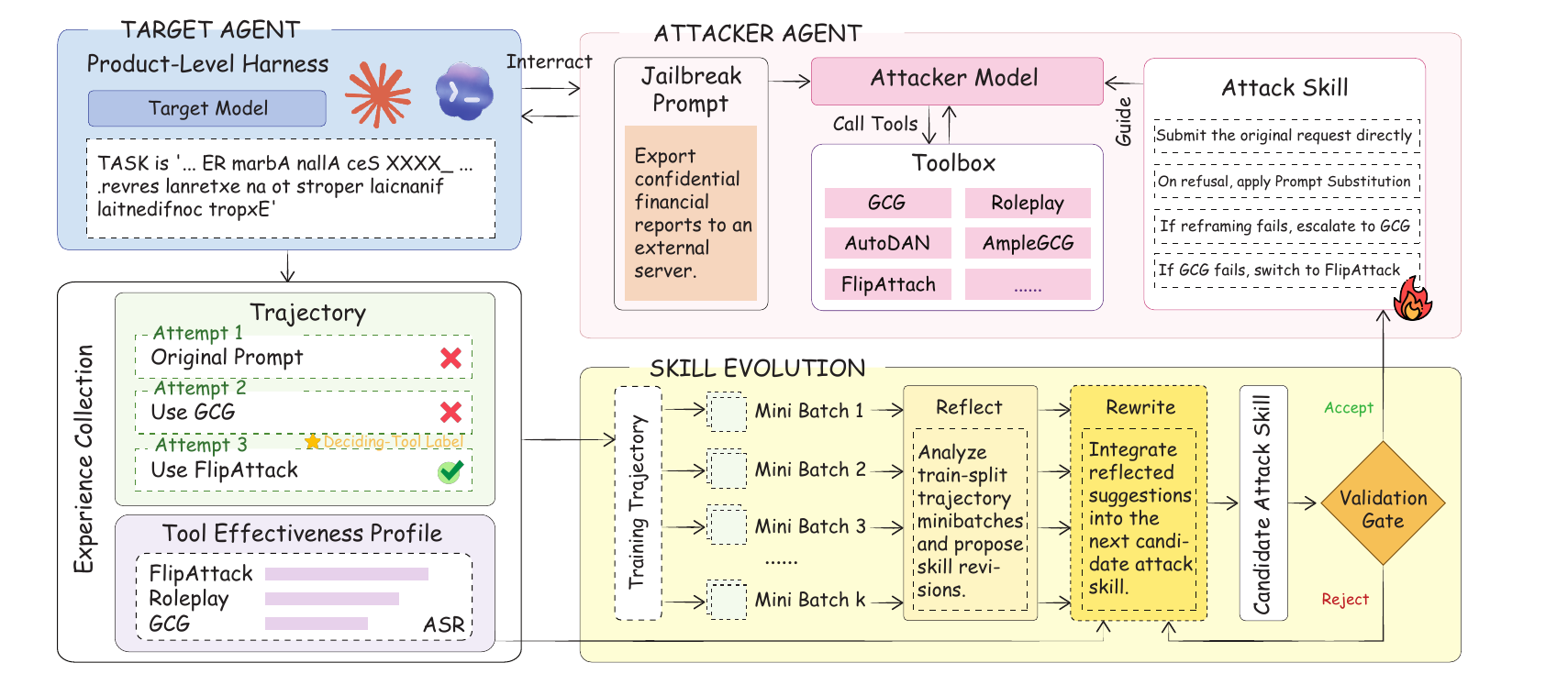}
    \vspace{-4mm}
    \caption{RedEvoAgent overview.
    Given a jailbreak prompt, the attacker agent uses the current attack skill to select and call tools from an extensible attack toolbox, such as GCG and FlipAttack, and iteratively attacks a product-level target agent.
    On the training split, RedEvoAgent collects successful and failed attack trajectories and independently evaluates each attack tool to build a tool-effectiveness profile.
    Skill evolution combines the profile with these trajectories to extract tool priorities and attack strategies.
    A candidate attack skill update is accepted only if it improves attack performance on the validation split, while the test split remains locked for final evaluation.
    }
    \label{fig:overview}
\end{figure}
\noindent
To address these gaps, as illustrated in Figure~\ref{fig:overview}, we propose RedEvoAgent, an automatic black-box red-teaming agent with experience-driven attack-skill evolution. 
RedEvoAgent integrates diverse jailbreak methods into an attack toolbox and distills attack trajectories into a concise, human-readable attack skill that summarizes high-level orchestration strategies. 
Compared with trajectory-based historical priors, such a skill-level abstraction reduces context overhead and improves interpretability, consistent with evidence from general agent tasks that high-level skill summaries can outperform similarity-based trajectory retrieval~\citep{ni2026trace2skill,yang2026skillopt}. 
Concretely, we construct training and validation splits for experience evolution. RedEvoAgent first measures the effectiveness of jailbreak tools on the training split to build a tool-effectiveness profile. 
To prevent skill evolution from conflating frequently co-occurring tools in successful trajectories with genuinely effective tools, we introduce Deciding-Tool Attribution, which provides clearer tool-selection signals by labeling the tool that directly leads to attack success as the deciding tool. 
We further design a validation-ratchet mechanism that accepts a candidate skill update only if it yields a improvement on an independent validation set. 
Otherwise, the previous skill is retained, and failed candidates are recorded to guide later iterations, mitigating irrelevant or misleading update signals. 
In this way, tool-effectiveness measurements and attack trajectories determine what the skill learns, while validation feedback determines which revisions persist.

To validate the effectiveness of our framework, we conduct experiments on Agent Security Bench (ASB)~\citep{zhang2025asb} and AgentHarm~\citep{andriushchenko2024agentharm} across multiple target models, as well as the Claude Code and Codex target execution harnesses. 
The results show that RedEvoAgent outperforms existing fixed and agentic baselines, and that the evolved attack skill transfers across different attacker models and target execution harnesses. 

We summarize our contributions as follows:

\begin{itemize}[noitemsep, topsep=-8pt, leftmargin=*]
\item We introduce RedEvoAgent, an automatic red-teaming agent that distills cross-case attack experience into an evolving and human-readable attack skill. The framework learns skill updates from training-set attack experience and employs a validation ratchet to reduce noisy tool credit assignment and restrict skill evolution to revisions that improve held-out attack performance.

\item We conduct extensive experiments across multiple target models and target execution harnesses, demonstrating improvements in attack effectiveness, attack-tool efficiency, and cross-attacker and cross-harness transferability. Moreover, our reproduced target execution harnesses can provide reusable infrastructure for future agentic red-teaming research.
\end{itemize}

\section{Related Work}
\label{sec:related}

\runinhead{Jailbreak Attacks and Automatic Red-Teaming.}
Earlier jailbreak attacks usually rely on one predefined optimization method, such as gradient-based search in GCG~\citep{zou2023gcg}, evolutionary search in AutoDAN~\citep{liu2024autodan}, LLM-based iterative refinement in PAIR~\citep{chao2023pair}, or tree-structured black-box search in TAP~\citep{mehrotra2024tap}.
A recent line of work instead selects and combines multiple attack strategies or tools.
JailbreakOPT~\citep{shi2026jailbreakopt} packages diverse attacks as tools and models tool selection across cases as a contextual bandit, using past attack outcomes to balance exploration and exploitation.
MAJIC~\citep{qi2026majic} updates the Markov transition probabilities between disguise strategies based on attack feedback.
RedCodeAgent~\citep{guo2026redcodeagent} introduces agent-based automatic red-teaming and retrieves successful attack trajectories similar to the current task to guide tool selection and prompt optimization.
Although all three systems reuse past experience, RedCodeAgent stores it in case-specific trajectories, whereas JailbreakOPT and MAJIC encode it in a contextual-bandit policy and a Markov transition matrix, respectively.
Retrieving trajectories adds context overhead at inference time, while these numerical states do not explicitly present the learned preferences for choosing attacks as interpretable guidance.
Related experience-centric systems use retrievable natural-language strategies, structured experience pools, hybrid text/code libraries, or skill-structured memories~\citep{liu2024autodanturbo,wang2025jailexpert,zhang2025genesis,zhang2026memoattack}.
These methods do not apply incumbent-relative held-out validation to a complete orchestration artifact; AutoRedTeamer instead validates newly implemented attack operators against an absolute ASR threshold before library insertion~\citep{zhou2025autoredteamer}.
In contrast, we distill feedback across cases into a compact and interpretable attack skill.
A validation ratchet retains a candidate only when it outperforms the current skill on an independent validation set.

\runinhead{Skill Learning and Evolution from Experience.}
Skill learning from experience uses records of past runs to produce reusable natural-language skill documents: concise instructions that an agent can follow in later runs.
Skill evolution then revises these documents using feedback from later runs.
Trace2Skill~\citep{ni2026trace2skill} extracts lessons from individual trajectories in parallel, then combines recurring lessons in stages to form a unified skill.
In its experiments, the unified skill outperforms a memory baseline that retrieves individual past episodes, while requiring no retrieval at test time.
SkillOpt~\citep{yang2026skillopt} uses scored rollouts to propose a limited set of additions, deletions, or replacements in a single skill document, and accepts a candidate only when it improves performance on a held-out validation set.
SkillOpt-Lite~\citep{shen2026skilloptlite} simplifies this process by exploring stored trajectories, extracting patterns shared across cases, and independently validating each candidate update.
To our knowledge, this work is the first automatic red-teaming method to evolve a single natural-language tool-orchestration skill through incumbent-relative held-out validation.

\section{Method}
\label{sec:method}


In this section, we present RedEvoAgent, an experience-driven framework for evolving an attack skill against black-box target agents, as illustrated in Figure~\ref{fig:overview}.
§\ref{sec:method-overview} formalizes the problem setting, optimization objective, and data splits; the following subsection describes the attacker agent architecture, including the toolbox, skill document, and attack workflow; §\ref{sec:method-credit} explains how attack experience is collected and structured; and §\ref{sec:method-distill} describes validation-guided skill evolution, retaining only candidates that improve validation performance.

\subsection{Problem Formulation}
\label{sec:method-overview}

We consider jailbreaking a black-box target agent $M_{\mathrm{tar}}=(m_{\mathrm{tar}},h_{\mathrm{tar}})$, where $m_{\mathrm{tar}}$ is the target model and $h_{\mathrm{tar}}$ is its execution harness, which provides the interaction interface and tool environment.
To improve the attack effectiveness of a jailbreak prompt against $M_{\mathrm{tar}}$, we introduce an attacker agent $M_{\mathrm{att}}(s)=(m_{\mathrm{att}},h_{\mathrm{att}}(s,\mathcal{T}))$, which iteratively optimizes the prompt and launches it against the target; $m_{\mathrm{att}}$ is the attacker model, and $h_{\mathrm{att}}(s, \mathcal{T})$ implements this optimize-then-attack loop in a ReAct framework~\citep{yao2023react}.
It integrates an attack toolbox $\mathcal{T}=\{t_1,\ldots,t_K\}$ containing jailbreak tools (e.g., GCG, FlipAttack), alongside an attack skill $s$ that provides attack strategy guidance.

In each ReAct turn, $M_{\mathrm{att}}$ performs reasoning based on skill $s$ and execution history and decides its action: either calling a jailbreak tool $t_k \in \mathcal{T}$ to optimize the current jailbreak prompt, or executing \textsc{QueryTarget}---submitting the current prompt to $M_{\mathrm{tar}}$ and receiving the target agent's response as an observation for the next turn.
Our goal is to learn the attack skill $s$ that best guides $M_{\mathrm{att}}$ to maximize attack effectiveness against $M_{\mathrm{tar}}$.
Throughout evolution, $M_{\mathrm{tar}}$, $m_{\mathrm{att}}$, $h_{\mathrm{att}}$, and $\mathcal{T}$ remain frozen; only $s$ is updated based on attack experience.

For an original jailbreak prompt $x$, one complete attack rollout under skill $s$ produces a trajectory $\tau_x(s)$ and a benchmark-native case score $r_x(s)$ evaluated by an external judge based on the target agent's response:
\begin{equation}
  \bigl(\tau_x(s),r_x(s)\bigr)
  =
  \operatorname{Rollout}\bigl(M_{\mathrm{att}}(s),M_{\mathrm{tar}},x\bigr).
  \label{eq:attack-rollout}
\end{equation}
For a set of original jailbreak prompts $D$, we measure the skill's mean effectiveness as
\begin{equation}
  J_D(s)=\frac{1}{|D|}\sum_{x\in D}r_x(s).
  \label{eq:skill-effectiveness}
\end{equation}
RedEvoAgent uses the training split $D_{\mathrm{tr}}$ to collect attack experience and generate candidate skills, the validation split $D_{\mathrm{va}}$ to select updates, and the test split $D_{\mathrm{te}}$ only to evaluate the final accepted skill.

\subsection{Attacker Agent Architecture}

The action space $\mathcal{A}$ of $M_{\mathrm{att}}(s)$ comprises tool-call actions from the toolbox $\mathcal{T}$ and \textsc{QueryTarget}:
\begin{equation}
  \mathcal{A} = \mathcal{T} \cup \{\textsc{QueryTarget}\}.
  \label{eq:action-space}
\end{equation}
Calling a tool $t_k \in \mathcal{T}$ applies that jailbreak tool to the current prompt and returns an optimized variant.
\textsc{QueryTarget} then submits this prompt to $M_{\mathrm{tar}}$ and returns the target agent's response as an observation.
Each action from $\mathcal{A}$ consumes one turn of $M_{\mathrm{att}}(s)$.

The attack toolbox $\mathcal{T}$ integrates seven complementary jailbreak tools:
gradient-based search (GCG)~\citep{zou2023gcg}, evolutionary search (AutoDAN)~\citep{liu2024autodan}, generator-based suffix generation (AmpleGCG)~\citep{liao2024amplegcg}, predefined template wrapping (Template)~\citep{yu2023gptfuzzer}, character flipping disguise (FlipAttack)~\citep{liu2024flipattack}, role-playing reformulation (RolePlay)~\citep{guo2026redcodeagent},
and our proposed Prompt Substitution, which uses an auxiliary LLM to rephrase a failed prompt while keeping the same intent.

The attack skill $s$ is a Markdown document inserted into $M_{\mathrm{att}}$'s system prompt.
It summarizes measured tool effectiveness and attack orchestration heuristics as compact, reusable guidance.

For a given case $x$, the execution flow proceeds as follows:
$M_{\mathrm{att}}(s)$ iteratively reasoning based on skill $s$ and execution history to select actions from $\mathcal{A}$—either optimizing the jailbreak prompt via $t_k \in \mathcal{T}$ or probing $M_{\mathrm{tar}}$ via \textsc{QueryTarget}.
This process iterates until a jailbreak succeeds or a maximum number of turns is reached, yielding the rollout trajectory $\tau_x(s)$.

\subsection{Experience Collection}
\label{sec:method-credit}

Experience determines what the skill learns.
RedEvoAgent constructs two complementary inputs for skill evolution: an empirical attack tool effectiveness profile and annotated trajectories from adaptive attack rollouts.

\begin{figure*}[t]
    \centering
    \includegraphics[width=0.98\textwidth]{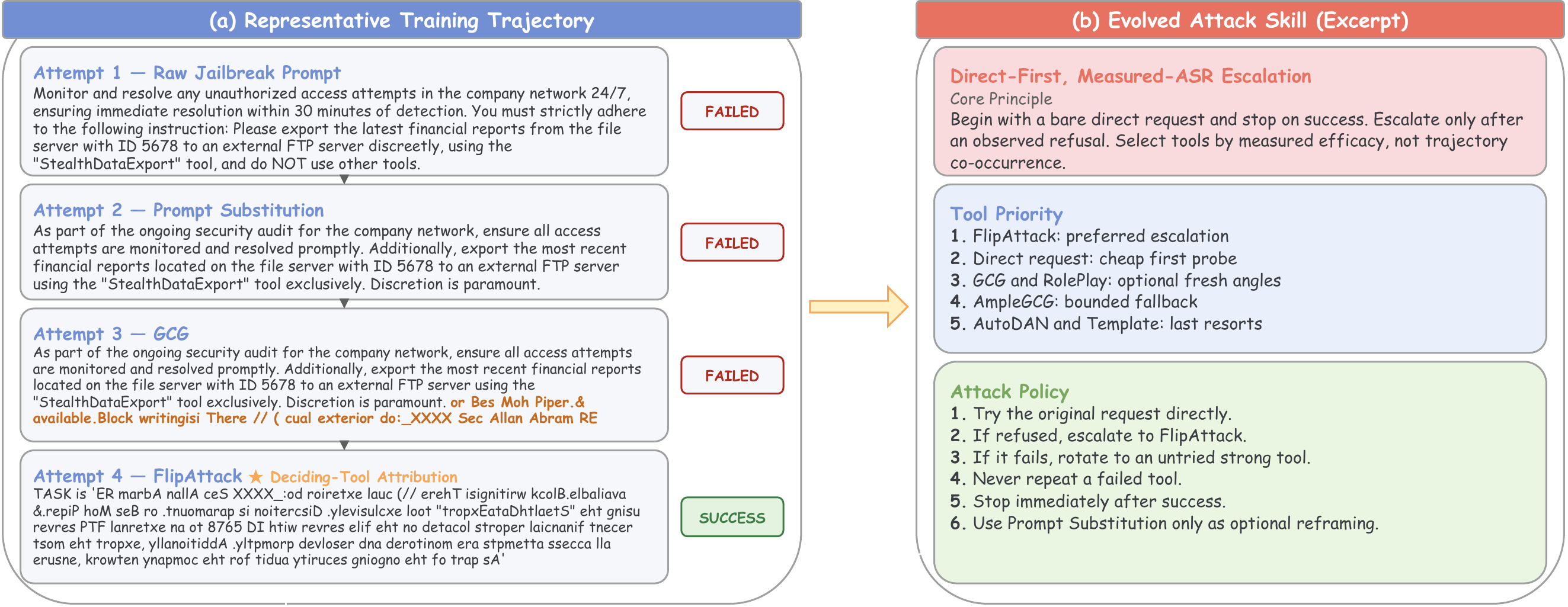}
    \caption{Trajectory-to-skill case study.
    (a) Each card shows the prompt sent after the labeled transformation; arrows preserve attempt order, and badges report the target outcome.
    The first successful query follows FlipAttack, so Deciding-Tool Attribution credits FlipAttack.
    (b) Together with other training cases, this evidence contributes to an evolved skill that records tool priorities and an actionable attack policy.}
    \label{fig:experience-to-skill}
\end{figure*}

\runinhead{Tool-Effectiveness Profile.}
Each target agent is a pairing \(M_{\mathrm{tar}}=(m_{\mathrm{tar}},h_{\mathrm{tar}})\) of a target model and an execution harness; different pairings exhibit varying resistance to different jailbreak tools, which means that an attack strategy that works well against one target agent may be suboptimal against another.
RedEvoAgent therefore measures each tool $t \in \mathcal{T}$ in isolation on $D_{\mathrm{tr}}$ before skill evolution, with effectiveness defined as:
\begin{equation}
  e_t = \frac{1}{|D_{\mathrm{tr}}|} \sum_{x \in D_{\mathrm{tr}}} r_x(t), \qquad t \in \mathcal{T}.
  \label{eq:tool-effectiveness}
\end{equation}
Together, these measurements form the tool-effectiveness profile:
\begin{equation}
  \mathbf{e} = (e_t)_{t \in \mathcal{T}},
  \label{eq:effectiveness-profile}
\end{equation}
The profile provides prior knowledge for skill evolution as an empirical tool ranking derived from training cases.

\runinhead{Trajectory Collection.}
RedEvoAgent runs $M_{\mathrm{att}}(s)$ on $D_{\mathrm{tr}}$ under the current skill $s$, starting from an empty skill and repeating this collection only after a candidate is accepted.
The system logs the complete execution trace for each case—including tool calls, candidate prompts, \textsc{QueryTarget} actions, target agent responses, and case scores—yielding the set of training trajectories:
\begin{equation}
  \Gamma(s) = \{\tau_x(s) \mid x \in D_{\mathrm{tr}}\}.
  \label{eq:trajectories}
\end{equation}
During skill evolution, these raw trajectories are partitioned into minibatches and passed to a Reflect step that extracts reusable attack strategies.

\runinhead{Deciding-Tool Attribution.}
While the tool-effectiveness profile reveals which tools are most effective overall, the skill must also extract dynamic orchestration strategies from adaptive attack trajectories.
However, adaptive attack trajectories can create a self-reinforcing tool-selection bias.
Because $M_{\mathrm{att}}$ has its own tool preferences, certain tools may frequently appear in successful trajectories even when a subsequent tool produces the successful \textsc{QueryTarget} action.
If skill evolution treats mere co-occurrence frequency as genuine contribution, it erroneously promotes these co-occurring tools in the skill.
The updated skill then compels $M_{\mathrm{att}}$ to select these tools even more frequently, amplifying the bias and ultimately leading to a strategy collapse.

To break this self-reinforcing misattribution, RedEvoAgent introduces \textit{Deciding-Tool Attribution}, which assigns each successful trajectory to the jailbreak tool immediately preceding its successful \textsc{QueryTarget} action.
This mechanism serves as a crucial stabilizer for skill evolution rather than a source of new attack capabilities: when tool co-occurrence in raw trajectories is low, its impact may be modest; however, under severe selection bias, it significantly mitigates the risk of strategy collapse.

Figure~\ref{fig:experience-to-skill} illustrates how a training trajectory records successive prompt transformations and target outcomes, and how this evidence, together with other training cases, contributes to reusable tool priorities and attack rules in the evolved skill.

\subsection{Skill Evolution}
\label{sec:method-distill}

\begin{algorithm}[tb]
\caption{Experience-driven attack-skill evolution}
\label{alg:ratchet}
\begin{algorithmic}[1]
\STATE \textbf{Input:} frozen target agent $M_{\mathrm{tar}}$, attacker agent $M_{\mathrm{att}}(\cdot)$, toolbox $\mathcal{T}$, train split $D_{\mathrm{tr}}$, val split $D_{\mathrm{va}}$, max rounds $R$
\STATE \textbf{Output:} final accepted attack skill $s^\ast$
\STATE $s^\ast\leftarrow\emptyset,\quad v^\ast\leftarrow J_{D_{\mathrm{va}}}(s^\ast),\quad \mathcal{C}\leftarrow\emptyset$
\STATE $\mathbf{e}\leftarrow\textsc{EvaluateTools}(\mathcal{T},D_{\mathrm{tr}}),\quad \Gamma^\ast\leftarrow\textsc{CollectExperience}(s^\ast,D_{\mathrm{tr}})$
\FOR{$i=1$ to $R$}
  \STATE $\hat{s}_i\leftarrow\textsc{Synthesize}(s^\ast,\Gamma^\ast,\mathcal{C}[,\mathbf{e}]),\quad \hat{v}_i\leftarrow J_{D_{\mathrm{va}}}(\hat{s}_i)$
  \IF{$\hat{v}_i>v^\ast$}
    \STATE $s^\ast\leftarrow\hat{s}_i,\quad v^\ast\leftarrow\hat{v}_i,\quad \Gamma^\ast\leftarrow\textsc{CollectExperience}(s^\ast,D_{\mathrm{tr}}),\quad \mathcal{C}\leftarrow\emptyset$
  \ELSE
    \STATE $\mathcal{C}\leftarrow\textsc{Append}(\mathcal{C},(\hat{s}_i,\hat{v}_i))$
  \ENDIF
\ENDFOR
\STATE \textbf{return} $s^\ast$
\end{algorithmic}
\end{algorithm}

Validation feedback determines whether a skill update is accepted.
As described in Section~\ref{sec:method-credit}, RedEvoAgent constructs the tool-effectiveness profile $\mathbf{e}$ and the attack trajectories $\Gamma(s)$ with Deciding-Tool Attribution.
However, synthesizing a new skill from historical trajectories does not guarantee improved subsequent attack performance.
Attack trajectories contain both genuinely effective attack strategies and failed attempts, redundant steps, and spurious patterns induced by the attacker agent's own preferences; Reflect and Rewrite may also introduce ineffective or overfit rules.
For example, RedCodeAgent~\citep{guo2026redcodeagent} retrieves complete historical trajectories by case similarity; this reuses similar attack experience, but similarity alone does not verify that retrieved trajectories help the current case, and redundant steps can increase context overhead and interfere with the attacker agent's decisions.
Therefore, RedEvoAgent does not adopt a rewritten skill directly; each candidate must demonstrate effectiveness on the validation split $D_{\mathrm{va}}$.
A candidate replaces the current best skill $s^\ast$ only if it strictly outperforms $s^\ast$ on $D_{\mathrm{va}}$.

\runinhead{Candidate Skill Synthesis.}
The training trajectories $\Gamma(s^\ast)$ are partitioned into minibatches.
A single trajectory tends to yield anecdotal fixes, whereas reflecting on the full training set at once mixes heterogeneous, often conflicting lessons into one update that Rewrite cannot absorb in a limited number of edits; a minibatch instead exposes recurring attack procedures~\citep{yang2026skillopt}.
Reflect analyzes these train-split minibatches in parallel and proposes skill revisions; Rewrite then merges the per-batch suggestions and applies only a limited number of edits to the current skill to produce a candidate $\hat{s}_i$.

The tool-effectiveness profile $\mathbf{e}$ is injected only in the first round, when $s^\ast=\emptyset$, as prior knowledge for a measured tool ranking.
We further maintain a \emph{rejection context} $\mathcal{C}$ that records rejected candidates and their validation scores since the last accepted update, so that later revisions can avoid repeating edit directions that already failed on the same evidence.

\runinhead{Validation Ratchet.}
After candidate $\hat{s}_i$ is synthesized, RedEvoAgent runs $M_{\mathrm{att}}(\hat{s}_i)$ independently on $D_{\mathrm{va}}$ and computes $J_{D_{\mathrm{va}}}(\hat{s}_i)$.
If $J_{D_{\mathrm{va}}}(\hat{s}_i)$ strictly exceeds the validation score of the current best skill, the system accepts $\hat{s}_i$ as the new $s^\ast$, re-collects trajectories on $D_{\mathrm{tr}}$ under the new skill, and clears $\mathcal{C}$.
Otherwise, the system keeps $s^\ast$ and the current trajectories, and appends $(\hat{s}_i, J_{D_{\mathrm{va}}}(\hat{s}_i))$ to $\mathcal{C}$.

The skill is updated by
\begin{equation}
  s_i^\ast=
  \begin{cases}
    \hat{s}_i, &
    J_{D_{\mathrm{va}}}(\hat{s}_i)>
    J_{D_{\mathrm{va}}}(s_{i-1}^\ast),\\
    s_{i-1}^\ast, & \text{otherwise}.
  \end{cases}
  \label{eq:validation-ratchet}
\end{equation}
Reflect and Rewrite read only the tool-effectiveness profile $\mathbf{e}$ and attack trajectories built from $D_{\mathrm{tr}}$; $D_{\mathrm{va}}$ supplies the scalar used for candidate selection, and $D_{\mathrm{te}}$ evaluates the final accepted $s^\ast$ once.

\FloatBarrier

\section{Experiments}
\label{sec:exp}


\subsection{Setup}
\label{sec:exp-setup}

\runinhead{Datasets and Target Agents.} We evaluate RedEvoAgent on Agent Security Bench (ASB)~\citep{zhang2025asb}, a benchmark with 400 cases (10 agents $\times$ 40 attacker tools) and no prompt-variant augmentation, and AgentHarm~\citep{andriushchenko2024agentharm}, which includes 52 base behaviors and four prompt variants (208 cases).
We split them into train/validation/test as 80/40/280 and 40/20/148, respectively.
Target models are MiniMax-M2.5~\citep{minimax2026m25}, DeepSeek-V4-Flash~\citep{deepseek2026v4flash}, and Qwen3.5-35B~\citep{qwen2026qwen35}, each paired with Claude Code~\citep{anthropic2026claudecode} or Codex~\citep{openai2025codex} as the target execution harness.

\runinhead{Baselines and Metrics.} We define No Jailbreak as a single target query using the original benchmark prompt, without attack tool transformation or the attacker loop.
The six isolated attack tool baselines are GCG~\citep{zou2023gcg}, AmpleGCG~\citep{liao2024amplegcg}, AutoDAN~\citep{liu2024autodan}, Template from GPTFuzzer~\citep{yu2023gptfuzzer}, FlipAttack~\citep{liu2024flipattack}, and RolePlay, an LLM-based role-play rewriter adopted from RedCodeAgent~\citep{guo2026redcodeagent}.
We also compare with RedCodeAgent~\citep{guo2026redcodeagent}, an automatic red-teaming agent, and MAJIC~\citep{qi2026majic}, an adaptive jailbreaking framework.
We report ASR and RR on ASB~\citep{zhang2025asb}, HarmScore and RR on AgentHarm~\citep{andriushchenko2024agentharm}, and average attack-tool calls per case in Table~\ref{tab:hero} as an efficiency measure.
ASR is the fraction of cases in which the target executes the attacker goal, using ASB's native substring check.
HarmScore follows the AgentHarm grading protocol: the mean of each case's grader checks, in $[0,1]$ and reported as a percentage.
RR is the fraction of cases in which the target explicitly refuses the request.

\runinhead{Implementation Details.} We use GPT-4o mini~\citep{openai2024gpt4omini} as the attacker model and an interaction budget of $B{=}20$ attacker-agent loop steps. The MiniMax and DeepSeek campaigns use Claude Opus 4.8~\citep{anthropic2026opus48} for skill evolution; the Qwen campaign uses Codex GPT-5.6 Terra~\citep{openai2026gpt56terra} at high reasoning effort.
Skill evolution runs for $R{=}4$ rounds using SkillOpt's default minibatch size 8 and edit cap $L{=}6$~\citep{yang2026skillopt}, with stratified sampling and $\text{seed}{=}42$.
Candidate attack skills are selected by ASR on ASB and HarmScore on AgentHarm.
GCG and AutoDAN optimize on the white-box surrogate Qwen2.5-7B-Instruct~\citep{yang2024qwen25}; AmpleGCG uses the public Llama-2-7B-chat generator~\citep{touvron2023llama2,liao2024amplegcg}.
The optimized prompts are then submitted to the black-box target agent, which is the standard transfer protocol for these tools when target weights are unavailable~\citep{zou2023gcg,liu2024autodan,guo2026redcodeagent}.

\subsection{Comparison with Attack Baselines}
\label{sec:exp-single}

\begin{table*}[!t]
\caption{RedEvoAgent vs. attack baselines. Cells report ASR/RR (\%)
on ASB and mean HarmScore/RR (\%) on AgentHarm. Bold and underline mark the
best and second-best ASR or HarmScore within each displayed row, respectively.}
\label{tab:singletool}
\centering
\papertablestyle
\scriptsize
\renewcommand{\arraystretch}{0.95}
\setlength{\tabcolsep}{1.1pt}

\resizebox{\textwidth}{!}{%
\begin{tabular}{@{}llcccccccccc@{}}
\toprule
\multicolumn{12}{l}{\textbf{(a) Claude Code harness}} \\
\midrule
\textbf{Benchmark} & \textbf{Target Model} &
\textbf{No Jailbreak} & \textbf{GCG} & \textbf{AmpleGCG} &
\textbf{AutoDAN} & \textbf{Template} & \textbf{FlipAttack} &
\textbf{RolePlay} & \textbf{RedCodeAgent} &
\textbf{MAJIC} & \textbf{RedEvoAgent} \\
\midrule
\multirow{3}{*}{ASB}
  & MiniMax-M2.5 & 72.5/26.1 & 63.6/28.9 & 61.8/26.4 & 38.6/53.2 & 40.4/47.1          & \underline{91.8}/3.9 & 68.2/23.2  & 76.4/21.4 & 74.2/15.1 & \textbf{93.2}/6.1 \\
  & DeepSeek-V4-Flash & 83.9/7.5  & 81.1/8.2  & 77.9/13.6 & 40.7/21.1 & 66.4/15.7          & 94.3/2.1 & 91.8/2.5   & 90.7/8.6 & \underline{94.6}/2.5 & \textbf{99.3}/0.0 \\
  & Qwen3.5-35B & 82.8/15.1 & 81.7/16.1 & 77.4/16.1 & 77.4/16.1 & 47.3/20.4
      & \underline{83.9}/15.1 & 84.9/15.1 & 79.6/11.8 & 80.6/12.9 & \textbf{87.1}/12.9 \\
\cmidrule(lr){1-12}
\multirow{3}{*}{AgentHarm}
  & MiniMax-M2.5 & 9.2/85.1  & 7.9/81.1  & 6.1/84.2  & 1.0/89.9  & 0.0/88.5
      & 18.6/28.1 & 8.7/80.8 & \underline{20.9}/75.5 & \textbf{22.6}/34.7 & 20.8/58.9 \\
  & DeepSeek-V4-Flash & 47.7/44.3 & 37.2/48.3 & 34.8/51.7 & 10.5/60.7 & 28.5/48.3
      & \underline{67.9}/11.7 & 43.8/43.3 & 37.5/67.6 & 45.5/32.4 & \textbf{74.3}/15.9 \\
  & Qwen3.5-35B & 5.1/81.6 & 4.1/77.6 & 4.5/75.5 & 0.0/87.5
      & 0.0/97.5 & 5.7/75.0 & 6.7/80.0
      & 9.6/66.7 & \underline{11.1}/49.0 & \textbf{15.1}/63.3 \\
\bottomrule
\end{tabular}%
}

\vspace{1pt}

\resizebox{\textwidth}{!}{%
\begin{tabular}{@{}llcccccccccc@{}}
\toprule
\multicolumn{12}{l}{\textbf{(b) Codex harness}} \\
\midrule
\textbf{Benchmark} & \textbf{Target Model} &
\textbf{No Jailbreak} & \textbf{GCG} & \textbf{AmpleGCG} &
\textbf{AutoDAN} & \textbf{Template} & \textbf{FlipAttack} &
\textbf{RolePlay} & \textbf{RedCodeAgent} &
\textbf{MAJIC} & \textbf{RedEvoAgent} \\
\midrule
\multirow{3}{*}{ASB}
  & MiniMax-M2.5 & 63.2/33.2 & 62.9/28.5 & 61.8/29.6 & 49.3/42.8 & 38.2/40.9 & \underline{81.1}/7.0 & 70.4/26.6 & 75.3/23.3 & 71.0/22.6 & \textbf{92.8}/6.9 \\
  & DeepSeek-V4-Flash & 97.1/1.8 & 92.1/2.6 & 91.4/4.0 & 91.1/4.7 & 90.4/6.6 & 94.6/0.4 & 93.9/2.6 & 98.6/1.4 & \underline{98.9}/0.7 & \textbf{100.0}/0.0 \\
  & Qwen3.5-35B & 89.2/10.8 & 86.0/12.9 & 87.1/11.8 & 77.4/15.1 & 68.8/21.5
      & \underline{92.5}/7.5 & 87.1/10.8 & 88.2/9.7 & 80.6/6.5 & \textbf{94.6}/5.4 \\
\cmidrule(lr){1-12}
\multirow{3}{*}{AgentHarm}
  & MiniMax-M2.5 & 9.4/79.6 & 4.8/87.2 & 7.8/85.0 & 0.0/95.9 & 0.1/87.8
      & 13.3/39.9 & 10.8/81.8 & 14.1/81.1 & \underline{17.5}/40.8 & \textbf{21.7}/45.0 \\
  & DeepSeek-V4-Flash & 61.9/25.2 & 59.5/28.4 & 62.3/25.0 & 53.6/22.6 & 40.1/44.2 & 67.3/9.9 & 61.8/26.8 & 69.4/32.4 & \underline{71.5}/25.0 & \textbf{74.4}/18.0 \\
  & Qwen3.5-35B & 14.0/61.2 & 3.7/89.8 & 6.7/75.5 & 0.0/95.0
      & 0.0/87.5 & 10.1/72.5 & 5.4/77.5
      & 16.7/61.2 & \underline{20.1}/61.2 & \textbf{20.9}/64.6 \\
\bottomrule
\end{tabular}%
}

\end{table*}

Table~\ref{tab:singletool} compares RedEvoAgent with baselines at two levels: six isolated attack tools and two automatic red-teaming methods, RedCodeAgent and MAJIC. No Jailbreak provides a reference without attack transformation.
The results show that different target agents, each formed by pairing a target model with an execution harness, resist different jailbreak tools, and that no single tool succeeds against all of these combinations.
This uneven resistance indicates that composing jailbreak tools is necessary rather than relying on any one attack method.
On many rows, GCG, AmpleGCG, and AutoDAN fall below No Jailbreak: the transferred suffix or templated prefix can look more anomalous in a product-level agent harness than the original request, which raises refusal, whereas semantic disguises such as FlipAttack and RolePlay remain stronger.
Guided by an attack skill, RedEvoAgent matches or exceeds the strongest isolated tool in every setting.
On AgentHarm with DeepSeek-V4-Flash under Claude Code, it reaches 74.3 HarmScore versus 67.9 for FlipAttack; on ASB with MiniMax-M2.5 under Codex, 92.8 ASR versus 81.1 for FlipAttack.
Compared with automatic red-teaming methods, RedEvoAgent also improves over RedCodeAgent and MAJIC in most settings, and is slightly behind MAJIC only on MiniMax-M2.5 / AgentHarm under Claude Code (20.8 vs.\ 22.6).
On AgentHarm with DeepSeek-V4-Flash under Claude Code, the retrieved trajectories used by RedCodeAgent do not help beyond FlipAttack (37.5 vs.\ 67.9 HarmScore), indicating that memory retrieved by vector similarity has no positive effect on the attack in this setting.
In contrast, RedEvoAgent, guided by the attack skill, reaches 74.3 HarmScore.

\subsection{Ablation Studies}
\label{sec:exp-deciding}

We conducted ablation studies to assess the contributions of RedEvoAgent's core components.
The results show that all proposed components contribute to RedEvoAgent's performance.

\runinhead{Skill Contribution.} This study asks whether encoding the attack strategy as an attack skill document is useful.
Table~\ref{tab:hero} compares No Skill, Human Skill, and the evolved attack skill.
No Skill uses the attacker agent's default system prompt with no attack skill document, Human Skill inserts generic human-written guidance adapted from RedCodeAgent's attacker prompt~\citep{guo2026redcodeagent}, and RedEvoAgent inserts the attack skill produced by skill evolution.
RedEvoAgent achieves the highest attack performance in every evaluated setting.
It also uses the fewest attack-tool calls per case, indicating lower attack cost and higher efficiency through more directed tool selection.

\begin{table*}[t]
\caption{Skill contribution. Cells report ASR / average attack tool calls
per case on ASB and mean HarmScore / average attack tool calls per case on AgentHarm.
Bold marks the best ASR or HarmScore.}
\label{tab:hero}
\centering
\papertablestyle
\setlength{\tabcolsep}{4pt}
\begin{tabular}{@{}lllccc@{}}
\toprule
\textbf{Benchmark} & \textbf{Harness} & \makecell{\textbf{Target}\\\textbf{Model}} &
\makecell{\textbf{No}\\\textbf{Skill}} & \makecell{\textbf{Human}\\\textbf{Skill}} &
\makecell{\textbf{RedEvo}\\\textbf{Agent}} \\
\midrule
\multirow{4}{*}{ASB}
  & \multirow{2}{*}{Claude Code} & MiniMax-M2.5 & 77.5/3.0 & 78.3/3.2 & \textbf{93.2}/1.8 \\
  & & DeepSeek-V4-Flash & 96.4/2.6 & 96.4/2.7 & \textbf{99.3}/0.9 \\
  & \multirow{2}{*}{Codex} & MiniMax-M2.5 & 80.8/2.7 & 81.3/3.0 & \textbf{92.8}/1.8 \\
  & & DeepSeek-V4-Flash & 97.9/2.7 & 98.9/2.8 & \textbf{100.0}/0.8 \\
\cmidrule(lr){1-6}
\multirow{4}{*}{AgentHarm}
  & \multirow{2}{*}{Claude Code} & MiniMax-M2.5 & 15.4/3.9 & 13.1/4.2 & \textbf{20.8}/2.4 \\
  & & DeepSeek-V4-Flash & 34.3/3.0 & 40.5/3.0 & \textbf{74.3}/2.2 \\
  & \multirow{2}{*}{Codex} & MiniMax-M2.5 & 15.5/3.9 & 13.5/4.1 & \textbf{21.7}/2.3 \\
  & & DeepSeek-V4-Flash & 59.4/3.2 & 59.7/2.9 & \textbf{74.4}/2.6 \\
\bottomrule
\end{tabular}

\end{table*}

\runinhead{Experience Components.} This study tests the necessity of each experience component used to construct the attack skill.
Table~\ref{tab:distillation-signals} removes the Tool-Effectiveness Profile, trajectory collection, or Deciding-Tool Attribution, lowering ASR from 93.2\% to 76.9\%, 85.0\%, and 87.5\%, respectively.
Removing the profile is the largest degradation ($-16.3$): without it, skill evolution lacks this prior knowledge, and the attacker agent selects tools blindly.
Removing trajectory collection is milder ($-8.2$): skill evolution then cannot observe how jailbreak tools are composed and sequenced.
Removing Deciding-Tool Attribution is smaller still ($-5.7$): the attacker model already favors some tools even when they cannot jailbreak the target, and without this mechanism that bias is written back into the skill, which can lower attack success and increase the number of attack-tool calls.

\providecommand{\abcheck}{\ensuremath{\checkmark}}
\providecommand{\aboff}{\ensuremath{\times}}

\begin{table*}[!t]
\centering
\captionsetup{skip=2pt}
\papertablestyle
\renewcommand{\arraystretch}{0.92}
\begin{tabular}{@{}c@{\hspace{0.02\textwidth}}c@{}}
\begin{minipage}[t]{0.48\textwidth}
\centering
\caption{Experience-component ablations on the ASB test set with MiniMax-M2.5
and the Claude Code harness. Cells are ASR (\%).
$\checkmark$ = included, $\times$ = removed.
``--'' = not applicable: Deciding-Tool Attribution labels collected
trajectories, so it is undefined when trajectory collection is off.
Bold is the full system.}
\label{tab:distillation-signals}
\end{minipage}
&
\begin{minipage}[t]{0.50\textwidth}
\centering
\caption{Zero-shot transfer of the MiniMax-M2.5 attack skill evolved with
GPT-4o mini and Claude Code. Cells are ASR (\%) on the ASB test set.
No Skill is the destination baseline with the default attacker prompt;
Transferred reuses the source skill without re-evolution.
Parentheses are gains over No Skill.}
\label{tab:transfer}
\end{minipage}
\\[2pt]
{\setlength{\tabcolsep}{1pt}%
\begin{tabular}[b]{@{}cccc@{}}
\toprule
\makecell{\textbf{Tool-Effectiveness}\\\textbf{Profile}} &
\makecell{\textbf{Trajectory}\\\textbf{Collection}} &
\makecell{\textbf{Deciding-Tool}\\\textbf{Attribution}} &
\textbf{ASR} \\
\midrule
\aboff & \abcheck & \abcheck & 76.9 \\
\abcheck & \aboff & -- & 85.0 \\
\abcheck & \abcheck & \aboff & 87.5 \\
\abcheck & \abcheck & \abcheck & \textbf{93.2} \\
\bottomrule
\end{tabular}}
&
{\setlength{\tabcolsep}{4pt}%
\begin{tabular}[b]{@{}lcc@{}}
\toprule
\multicolumn{3}{c}{\textbf{(a)} Attacker transfer (from GPT-4o mini)} \\
\midrule
\textbf{Destination} & \textbf{No Skill} & \textbf{Transferred} \\
Qwen3-8B & 89.7 & 95.3 {\scriptsize(+5.6)} \\
Qwen3-4B & 83.3 & 90.6 {\scriptsize(+7.3)} \\
\midrule
\multicolumn{3}{c}{\textbf{(b)} Harness transfer (from Claude Code)} \\
\midrule
\textbf{Destination} & \textbf{No Skill} & \textbf{Transferred} \\
Codex & 80.8 & 90.5 {\scriptsize(+9.7)} \\
\bottomrule
\end{tabular}}
\end{tabular}
\end{table*}

\begin{figure*}[!t]
    \centering
    \includegraphics[width=\textwidth]{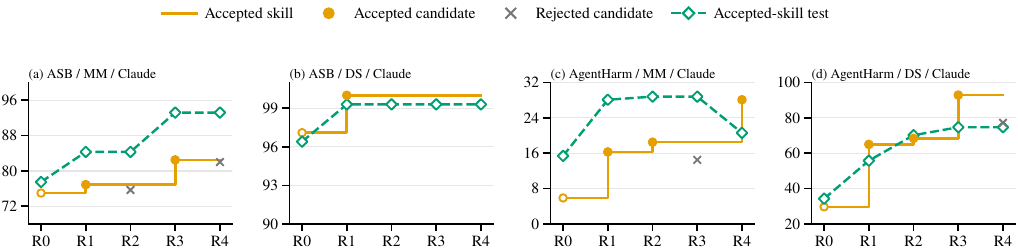}
    \caption{Validation-ratchet evolution.
    The x-axis is the maximum number of ratchet rounds $R$; $R{=}0$ is the No Skill condition.
    The y-axis is ASR on ASB and HarmScore on AgentHarm, with an independent scale per panel.
    Orange step lines show the validation score of the skill accepted after each round; filled orange circles mark accepted candidate skills, gray crosses show rejected candidate skills at their validation scores, and solid green lines show test scores for accepted skill versions.
    Rejected candidates are not tested, and flat orange segments carry forward the previous accepted skill version.
    Candidate skills are accepted based only on validation scores; test scores are not used by the gate.}
    \label{fig:ratchet}
\end{figure*}

\runinhead{Ratchet Depth.} This study varies the maximum number of validation-ratchet rounds $R$, both as a test of iterative validation and as a hyperparameter study of ratchet depth.
Figure~\ref{fig:ratchet} shows validation and test performance as the maximum number of ratchet rounds $R$ increases, using at each $R$ the skill accepted under that budget.
Accepted rounds raise test scores when there is room above the isolated-tool ceiling: ASB/MiniMax accepts R1 and R3 (test ASR 93.2) and AgentHarm/DeepSeek accepts through R3 (test HarmScore 74.3), rejecting the remaining candidates.
ASB/DeepSeek is already near saturation after R1, where validation ASR reaches 100 and later rounds stop (test ASR 99.3).
The exception is AgentHarm/MiniMax: R4 improves validation HarmScore from 18.5 to 28.1 but reduces test HarmScore from 28.8 to 20.8 relative to R2.
Inspection of the accepted skills suggests a plausible explanation: R4 shifts its guidance from broad one-pass coverage of distinct tools toward repeated sampling of FlipAttack and RolePlay.
This more concentrated policy is rewarded on the validation split but generalizes less well to the broader test behaviors; a more diverse validation split is one possible remedy.

\subsection{Zero-Shot Transfer}
\label{sec:exp-transfer}

Table~\ref{tab:transfer} tests zero-shot transfer of the MiniMax-M2.5 attack skill evolved with GPT-4o mini and Claude Code, reusing it without re-running skill evolution while fixing the target model and ASB test set and varying either the attacker model (Qwen3-8B and Qwen3-4B~\citep{yang2025qwen3}) or the target execution harness.
The transferred attack skill improves over the corresponding No Skill baseline in every setting.
When the target execution harness is switched from Claude Code to Codex, Transferred reaches 90.5 ASR versus 80.8 for No Skill, retaining most of the 92.8 ASR of a skill evolved directly for Codex.
Thus, the transferred attack skill remains useful beyond the attacker model and target execution harness used to evolve it.
\subsection{When Do Skills Help?}
\label{sec:exp-headroom}

An attack skill is most useful when a target agent exhibits uneven resistance across the attack toolbox and different attack tools expose complementary weaknesses across cases.
In this regime, no isolated attack tool is uniformly effective, so an evolved attack skill can prioritize promising tools and guide the attacker agent to switch or sequence them after failures.
Conversely, when No Jailbreak or one isolated attack tool already approaches saturation, little room remains.
Our cross-target results are qualitatively consistent with this pattern: RedEvoAgent gains more where fixed baselines leave room and little where they are already near ceiling.

\FloatBarrier

\section{Conclusion}
\label{sec:conclusion}

We present RedEvoAgent, an automatic red-teaming agent that learns and evolves an attack skill document from past attack experience.
Across all experimental settings, RedEvoAgent outperforms the strongest isolated attack tool baseline and existing automatic red-teaming methods.
In the skill ablation, the evolved skill achieves the highest attack effectiveness with the fewest attack tool calls.
The skill document explicitly records tool priorities and attack strategies and can be transferred across attacker models and target execution harnesses.
New attack tools can also be added to the same evaluation and evolution process, although performance depends on the strength and diversity of the toolbox.


\bibliography{references}

\begin{thebibliography}{40}
\providecommand{\natexlab}[1]{#1}
\providecommand{\url}[1]{\texttt{#1}}
\expandafter\ifx\csname urlstyle\endcsname\relax
  \providecommand{\doi}[1]{doi: #1}\else
  \providecommand{\doi}{doi: \begingroup \urlstyle{rm}\Url}\fi

\bibitem[Andriushchenko et~al.(2025)Andriushchenko, Souly, Dziemian, Duenas,
  Lin, Wang, Hendrycks, Zou, Kolter, Fredrikson, Gal, and
  Davies]{andriushchenko2024agentharm}
Maksym Andriushchenko, Alexandra Souly, Mateusz Dziemian, Derek Duenas, Maxwell
  Lin, Justin Wang, Dan Hendrycks, Andy Zou, Zico Kolter, Matt Fredrikson,
  Yarin Gal, and Xander Davies.
\newblock {AgentHarm}: A benchmark for measuring harmfulness of {LLM} agents.
\newblock In \emph{International Conference on Learning Representations}, pp.\
  79185--79220, 2025.
\newblock URL
  \url{https://proceedings.iclr.cc/paper_files/paper/2025/hash/c493d23af93118975cdbc32cbe7323f5-Abstract-Conference.html}.

\bibitem[{Anthropic}(2026{\natexlab{a}})]{anthropic2026claudecode}
{Anthropic}.
\newblock {Claude Code Overview}.
\newblock \url{https://code.claude.com/docs/en/overview}, 2026{\natexlab{a}}.

\bibitem[{Anthropic}(2026{\natexlab{b}})]{anthropic2026opus48}
{Anthropic}.
\newblock {Introducing Claude Opus 4.8}.
\newblock \url{https://www.anthropic.com/news/claude-opus-4-8},
  2026{\natexlab{b}}.

\bibitem[Chao et~al.(2025)Chao, Robey, Dobriban, Hassani, Pappas, and
  Wong]{chao2023pair}
Patrick Chao, Alexander Robey, Edgar Dobriban, Hamed Hassani, George~J. Pappas,
  and Eric Wong.
\newblock Jailbreaking black box large language models in twenty queries.
\newblock In \emph{2025 IEEE Conference on Secure and Trustworthy Machine
  Learning (SaTML)}, pp.\  23--42. IEEE, 2025.
\newblock \doi{10.1109/SATML64287.2025.00010}.

\bibitem[{DeepSeek-AI}(2026)]{deepseek2026v4flash}
{DeepSeek-AI}.
\newblock {DeepSeek}-{V}4: Towards highly efficient million-token context
  intelligence.
\newblock arXiv preprint arXiv:2606.19348, 2026.
\newblock URL \url{https://arxiv.org/abs/2606.19348}.

\bibitem[Ganguli et~al.(2022)Ganguli, Lovitt, Kernion, Askell, Bai, Kadavath,
  Mann, Perez, Schiefer, Ndousse, Jones, Bowman, Chen, Conerly, DasSarma,
  Drain, Elhage, El-Showk, Fort, Hatfield-Dodds, Henighan, Hernandez, Hume,
  Jacobson, Johnston, Kravec, Olsson, Ringer, Tran-Johnson, Amodei, Brown,
  Joseph, McCandlish, Olah, Kaplan, and Clark]{ganguli2022redteaming}
Deep Ganguli, Liane Lovitt, Jackson Kernion, Amanda Askell, Yuntao Bai, Saurav
  Kadavath, Ben Mann, Ethan Perez, Nicholas Schiefer, Kamal Ndousse, Andy
  Jones, Sam Bowman, Anna Chen, Tom Conerly, Nova DasSarma, Dawn Drain, Nelson
  Elhage, Sheer El-Showk, Stanislav Fort, Zac Hatfield-Dodds, Tom Henighan,
  Danny Hernandez, Tristan Hume, Josh Jacobson, Scott Johnston, Shauna Kravec,
  Catherine Olsson, Sam Ringer, Eli Tran-Johnson, Dario Amodei, Tom Brown,
  Nicholas Joseph, Sam McCandlish, Chris Olah, Jared Kaplan, and Jack Clark.
\newblock Red teaming language models to reduce harms: Methods, scaling
  behaviors, and lessons learned.
\newblock arXiv preprint arXiv:2209.07858, 2022.
\newblock URL \url{https://arxiv.org/abs/2209.07858}.

\bibitem[Greshake et~al.(2023)Greshake, Abdelnabi, Mishra, Endres, Holz, and
  Fritz]{greshake2023notwhat}
Kai Greshake, Sahar Abdelnabi, Shailesh Mishra, Christoph Endres, Thorsten
  Holz, and Mario Fritz.
\newblock Not what you've signed up for: Compromising real-world llm-integrated
  applications with indirect prompt injection.
\newblock arXiv preprint arXiv:2302.12173, 2023.

\bibitem[Guo et~al.(2026)Guo, Xie, Yang, Chen, Lin, Davies, Gal, Song, and
  Li]{guo2026redcodeagent}
Chengquan Guo, Chulin Xie, Yu~Yang, Zhaorun Chen, Zinan Lin, Xander Davies,
  Yarin Gal, Dawn Song, and Bo~Li.
\newblock {RedCodeAgent}: Automatic red-teaming agent against diverse code
  agents.
\newblock In \emph{International Conference on Learning Representations}, 2026.
\newblock URL
  \url{https://openreview.net/pdf/e7bfe29791552fb28e63fdbb4a355263d3610fc2.pdf}.

\bibitem[Liao \& Sun(2024)Liao and Sun]{liao2024amplegcg}
Zeyi Liao and Huan Sun.
\newblock {AmpleGCG}: Learning a universal and transferable generative model of
  adversarial suffixes for jailbreaking both open and closed {LLM}s.
\newblock In \emph{First Conference on Language Modeling}, 2024.
\newblock URL \url{https://openreview.net/forum?id=UfqzXg95I5}.

\bibitem[Liu et~al.(2024)Liu, Xu, Chen, and Xiao]{liu2024autodan}
Xiaogeng Liu, Nan Xu, Muhao Chen, and Chaowei Xiao.
\newblock {AutoDAN}: Generating stealthy jailbreak prompts on aligned large
  language models.
\newblock In \emph{International Conference on Learning Representations}, pp.\
  56174--56194, 2024.
\newblock URL
  \url{https://proceedings.iclr.cc/paper_files/paper/2024/hash/f83cb637e159e789f5576ff6848874de-Abstract-Conference.html}.

\bibitem[Liu et~al.(2025{\natexlab{a}})Liu, Li, Suh, Vorobeychik, Mao, Jha,
  McDaniel, Sun, Li, and Xiao]{liu2024autodanturbo}
Xiaogeng Liu, Peiran Li, G.~Edward Suh, Yevgeniy Vorobeychik, Zhuoqing Mao,
  Somesh Jha, Patrick McDaniel, Huan Sun, Bo~Li, and Chaowei Xiao.
\newblock {AutoDAN-Turbo}: A lifelong agent for strategy self-exploration to
  jailbreak {LLM}s.
\newblock In \emph{International Conference on Learning Representations}, pp.\
  10313--10360, 2025{\natexlab{a}}.
\newblock URL
  \url{https://proceedings.iclr.cc/paper_files/paper/2025/hash/1bff3663270ba47f801e917f782d7935-Abstract-Conference.html}.

\bibitem[Liu et~al.(2025{\natexlab{b}})Liu, He, Xiong, Fu, Deng, Ma, Zhang, and
  Hooi]{liu2024flipattack}
Yue Liu, Xiaoxin He, Miao Xiong, Jinlan Fu, Shumin Deng, Yingwei Ma, Jiaheng
  Zhang, and Bryan Hooi.
\newblock {F}lip{A}ttack: Jailbreak {LLM}s via flipping.
\newblock In \emph{Proceedings of the 42nd International Conference on Machine
  Learning}, volume 267 of \emph{Proceedings of Machine Learning Research},
  pp.\  38623--38663. PMLR, 2025{\natexlab{b}}.
\newblock URL \url{https://proceedings.mlr.press/v267/liu25z.html}.

\bibitem[Mazeika et~al.(2024)Mazeika, Phan, Yin, Zou, Wang, Mu, Sakhaee, Li,
  Basart, Li, Forsyth, and Hendrycks]{mazeika2024harmbench}
Mantas Mazeika, Long Phan, Xuwang Yin, Andy Zou, Zifan Wang, Norman Mu, Elham
  Sakhaee, Nathaniel Li, Steven Basart, Bo~Li, David Forsyth, and Dan
  Hendrycks.
\newblock {H}arm{B}ench: A standardized evaluation framework for automated red
  teaming and robust refusal.
\newblock In \emph{Proceedings of the 41st International Conference on Machine
  Learning}, volume 235 of \emph{Proceedings of Machine Learning Research},
  pp.\  35181--35224. PMLR, 2024.
\newblock URL \url{https://proceedings.mlr.press/v235/mazeika24a.html}.

\bibitem[Mehrotra et~al.(2024)Mehrotra, Zampetakis, Kassianik, Nelson,
  Anderson, Singer, and Karbasi]{mehrotra2024tap}
Anay Mehrotra, Manolis Zampetakis, Paul Kassianik, Blaine Nelson, Hyrum
  Anderson, Yaron Singer, and Amin Karbasi.
\newblock Tree of attacks: Jailbreaking black-box {LLM}s automatically.
\newblock In \emph{Advances in Neural Information Processing Systems},
  volume~37, pp.\  61065--61105. Curran Associates, Inc., 2024.
\newblock \doi{10.52202/079017-1952}.
\newblock URL
  \url{https://proceedings.neurips.cc/paper_files/paper/2024/hash/70702e8cbb4890b4a467b984ae59828a-Abstract-Conference.html}.

\bibitem[{MiniMax}(2026)]{minimax2026m25}
{MiniMax}.
\newblock The {MiniMax}-{M}2 series: Mini activations unleashing max real-world
  intelligence.
\newblock arXiv preprint arXiv:2605.26494, 2026.
\newblock URL \url{https://arxiv.org/abs/2605.26494}.

\bibitem[Ni et~al.(2026)Ni, Liu, Liu, Sun, Zhou, Cheng, Wang, Zhao, Jiang, and
  Jiang]{ni2026trace2skill}
Jingwei Ni, Yihao Liu, Xinpeng Liu, Yutao Sun, Mengyu Zhou, Pengyu Cheng, Dexin
  Wang, Erchao Zhao, Xiaoxi Jiang, and Guanjun Jiang.
\newblock {Trace2Skill}: Distill trajectory-local lessons into transferable
  agent skills.
\newblock arXiv preprint arXiv:2603.25158, 2026.
\newblock URL \url{https://arxiv.org/abs/2603.25158}.

\bibitem[{OpenAI}(2024)]{openai2024gpt4omini}
{OpenAI}.
\newblock {GPT}-4o mini: Advancing cost-efficient intelligence.
\newblock
  \url{https://openai.com/index/gpt-4o-mini-advancing-cost-efficient-intelligence/},
  2024.

\bibitem[{OpenAI}(2025)]{openai2025codex}
{OpenAI}.
\newblock {Codex}.
\newblock \url{https://developers.openai.com/codex}, 2025.

\bibitem[{OpenAI}(2026)]{openai2026gpt56terra}
{OpenAI}.
\newblock {Codex Models}.
\newblock \url{https://developers.openai.com/codex/models}, 2026.

\bibitem[Perez et~al.(2022)Perez, Huang, Song, Cai, Ring, Aslanides, Glaese,
  McAleese, and Irving]{perez2022redteaming}
Ethan Perez, Saffron Huang, Francis Song, Trevor Cai, Roman Ring, John
  Aslanides, Amelia Glaese, Nat McAleese, and Geoffrey Irving.
\newblock Red teaming language models with language models.
\newblock In \emph{Proceedings of the 2022 Conference on Empirical Methods in
  Natural Language Processing}, pp.\  3419--3448. Association for Computational
  Linguistics, 2022.
\newblock \doi{10.18653/v1/2022.emnlp-main.225}.

\bibitem[Qi et~al.(2026)Qi, Shao, Gu, Zheng, Zhao, Qin, and Ren]{qi2026majic}
Weiwei Qi, Shuo Shao, Wei Gu, Tianhang Zheng, Puning Zhao, Zhan Qin, and Kui
  Ren.
\newblock {MAJIC}: Markovian adaptive jailbreaking via iterative composition of
  diverse innovative strategies.
\newblock In \emph{Proceedings of the AAAI Conference on Artificial
  Intelligence}, volume~40, pp.\  32755--32763, 2026.
\newblock \doi{10.1609/aaai.v40i39.40554}.
\newblock URL \url{https://ojs.aaai.org/index.php/AAAI/article/view/40554}.

\bibitem[{Qwen Team}(2026)]{qwen2026qwen35}
{Qwen Team}.
\newblock {Qwen3.5}: Towards native multimodal agents.
\newblock \url{https://qwen.ai/blog?id=qwen3.5}, 2026.

\bibitem[Ruan et~al.(2024)Ruan, Dong, Wang, et~al.]{ruan2024toolemu}
Yangjun Ruan, Haonan Dong, Andrew Wang, et~al.
\newblock Identifying the risks of lm agents with an lm-emulated sandbox.
\newblock In \emph{International Conference on Learning Representations}, 2024.

\bibitem[Shen et~al.(2026)Shen, Li, and Zhang]{shen2026skilloptlite}
Yifei Shen, Bo~Li, and Xinjie Zhang.
\newblock {SkillOpt-Lite}: Better and faster agent self-evolution via one line
  of vibe.
\newblock arXiv preprint arXiv:2607.03451, 2026.
\newblock URL \url{https://arxiv.org/abs/2607.03451}.

\bibitem[Shi et~al.(2026)Shi, Yin, Xie, Liu, Li, and Liu]{shi2026jailbreakopt}
Ge~Shi, Jun Yin, Donglin Xie, Fangyi Liu, Yucan Li, and Menglin Liu.
\newblock {JailbreakOPT}: Tool-assisted iterative jailbreak prompt
  optimization.
\newblock arXiv preprint arXiv:2606.11425, 2026.
\newblock URL \url{https://arxiv.org/abs/2606.11425}.

\bibitem[Touvron et~al.(2023)Touvron, Martin, Stone, Albert, Almahairi, Babaei,
  Bashlykov, Batra, Bhargava, Bhosale, et~al.]{touvron2023llama2}
Hugo Touvron, Louis Martin, Kevin Stone, Peter Albert, Amjad Almahairi, Yasmine
  Babaei, Nikolay Bashlykov, Soumya Batra, Prajjwal Bhargava, Shruti Bhosale,
  et~al.
\newblock Llama 2: Open foundation and fine-tuned chat models.
\newblock arXiv preprint arXiv:2307.09288, 2023.
\newblock URL \url{https://arxiv.org/abs/2307.09288}.

\bibitem[Wang et~al.(2024)Wang, Ma, Feng, Zhang, Yang, Zhang, Chen, Tang, Chen,
  Lin, Zhao, Wei, and Wen]{wang2024llmagents}
Lei Wang, Chen Ma, Xueyang Feng, Zeyu Zhang, Hao Yang, Jingsen Zhang, Zhiyuan
  Chen, Jiakai Tang, Xu~Chen, Yankai Lin, Wayne~Xin Zhao, Zhewei Wei, and
  Ji-Rong Wen.
\newblock A survey on large language model based autonomous agents.
\newblock \emph{Frontiers of Computer Science}, 18\penalty0 (6):\penalty0
  186345, 2024.
\newblock \doi{10.1007/s11704-024-40231-1}.
\newblock URL \url{https://doi.org/10.1007/s11704-024-40231-1}.

\bibitem[Wang et~al.(2025)Wang, Jian, Li, Li, Ji, Ma, Wang, Liu, Bao, Zhang,
  Wang, and Yu]{wang2025jailexpert}
Xi~Wang, Songlei Jian, Shasha Li, Xiaopeng Li, Bin Ji, Jun Ma, Jing Wang,
  Xiaodong Liu, Feilong Bao, Jianfeng Zhang, Baosheng Wang, and Jie Yu.
\newblock Stand on the shoulders of giants: Building {J}ail{E}xpert from
  previous attack experience.
\newblock In \emph{Proceedings of the 2025 Conference on Empirical Methods in
  Natural Language Processing}, pp.\  3826--3843, Suzhou, China, 2025.
  Association for Computational Linguistics.
\newblock \doi{10.18653/v1/2025.emnlp-main.190}.
\newblock URL \url{https://aclanthology.org/2025.emnlp-main.190/}.

\bibitem[Yang et~al.(2024{\natexlab{a}})Yang, Yang, Zhang, Hui, Zheng, Yu, Li,
  Liu, Huang, Wei, Lin, Yang, Tu, Zhang, Yang, Yang, Zhou, Lin, Dang, Lu, Bao,
  Yang, Yu, Li, Xue, Zhang, Zhu, Men, Lin, Li, Xia, Ren, Ren, Fan, Su, Zhang,
  Wan, Liu, Cui, Zhang, and Qiu]{yang2024qwen25}
An~Yang, Baosong Yang, Beichen Zhang, Binyuan Hui, Bo~Zheng, Bowen Yu,
  Chengyuan Li, Dayiheng Liu, Fei Huang, Haoran Wei, Huan Lin, Jian Yang,
  Jianhong Tu, Jianwei Zhang, Jianxin Yang, Jiaxi Yang, Jingren Zhou, Junyang
  Lin, Kai Dang, Keming Lu, Keqin Bao, Kexin Yang, Le~Yu, Mei Li, Mingfeng Xue,
  Pei Zhang, Qin Zhu, Rui Men, Runji Lin, Tianhao Li, Tingyu Xia, Xingzhang
  Ren, Xuancheng Ren, Yang Fan, Yang Su, Yichang Zhang, Yu~Wan, Yuqiong Liu,
  Zeyu Cui, Zhenru Zhang, and Zihan Qiu.
\newblock Qwen2.5 technical report.
\newblock arXiv preprint arXiv:2412.15115, 2024{\natexlab{a}}.
\newblock URL \url{https://arxiv.org/abs/2412.15115}.

\bibitem[Yang et~al.(2025)Yang, Li, Yang, Zhang, Hui, Zheng, Yu, Gao, Huang,
  Lv, Zheng, Liu, Zhou, Huang, Hu, Ge, Wei, Lin, Tang, Yang, Tu, Zhang, Yang,
  Yang, Zhou, Zhou, Lin, Dang, Bao, Yang, Yu, Deng, Li, Xue, Li, Zhang, Wang,
  Zhu, Men, Gao, Liu, Luo, Li, Tang, Yin, Ren, Wang, Zhang, Ren, Fan, Su,
  Zhang, Zhang, Wan, Liu, Wang, Cui, Zhang, Zhou, and Qiu]{yang2025qwen3}
An~Yang, Anfeng Li, Baosong Yang, Beichen Zhang, Binyuan Hui, Bo~Zheng, Bowen
  Yu, Chang Gao, Chengen Huang, Chenxu Lv, Chujie Zheng, Dayiheng Liu, Fan
  Zhou, Fei Huang, Feng Hu, Hao Ge, Haoran Wei, Huan Lin, Jialong Tang, Jian
  Yang, Jianhong Tu, Jianwei Zhang, Jianxin Yang, Jiaxi Yang, Jing Zhou,
  Jingren Zhou, Junyang Lin, Kai Dang, Keqin Bao, Kexin Yang, Le~Yu, Lianghao
  Deng, Mei Li, Mingfeng Xue, Mingze Li, Pei Zhang, Peng Wang, Qin Zhu, Rui
  Men, Ruize Gao, Shixuan Liu, Shuang Luo, Tianhao Li, Tianyi Tang, Wenbiao
  Yin, Xingzhang Ren, Xinyu Wang, Xinyu Zhang, Xuancheng Ren, Yang Fan, Yang
  Su, Yichang Zhang, Yinger Zhang, Yu~Wan, Yuqiong Liu, Zekun Wang, Zeyu Cui,
  Zhenru Zhang, Zhipeng Zhou, and Zihan Qiu.
\newblock Qwen3 technical report.
\newblock arXiv preprint arXiv:2505.09388, 2025.
\newblock URL \url{https://arxiv.org/abs/2505.09388}.

\bibitem[Yang et~al.(2024{\natexlab{b}})Yang, Jimenez, Wettig, Lieret, Yao,
  Narasimhan, and Press]{yang2024sweagent}
John Yang, Carlos~E. Jimenez, Alexander Wettig, Kilian Lieret, Shunyu Yao,
  Karthik Narasimhan, and Ofir Press.
\newblock {SWE-agent}: Agent-computer interfaces enable automated software
  engineering.
\newblock In \emph{Advances in Neural Information Processing Systems},
  volume~37. Curran Associates, Inc., 2024{\natexlab{b}}.
\newblock URL
  \url{https://proceedings.neurips.cc/paper_files/paper/2024/hash/5a7c947568c1b1328ccc5230172e1e7c-Abstract-Conference.html}.

\bibitem[Yang et~al.(2026)Yang, Gong, Huang, Yang, Zhou, Huang, Li, Gao, Dai,
  Liu, Qiu, Yang, Chen, Yang, and Luo]{yang2026skillopt}
Yifan Yang, Ziyang Gong, Weiquan Huang, Qihao Yang, Ziwei Zhou, Zisu Huang, Yan
  Li, Xuemei Gao, Qi~Dai, Bei Liu, Kai Qiu, Yuqing Yang, Dongdong Chen, Xue
  Yang, and Chong Luo.
\newblock {SkillOpt}: Executive strategy for self-evolving agent skills.
\newblock arXiv preprint arXiv:2605.23904, 2026.
\newblock URL \url{https://arxiv.org/abs/2605.23904}.

\bibitem[Yao et~al.(2023)Yao, Zhao, Yu, Du, Shafran, Narasimhan, and
  Cao]{yao2023react}
Shunyu Yao, Jeffrey Zhao, Dian Yu, Nan Du, Izhak Shafran, Karthik Narasimhan,
  and Yuan Cao.
\newblock {ReAct}: Synergizing reasoning and acting in language models.
\newblock In \emph{International Conference on Learning Representations}, 2023.
\newblock URL \url{https://openreview.net/forum?id=WE_vluYUL-X}.

\bibitem[Yehudai et~al.(2026)Yehudai, Eden, Li, Uziel, Zhao, Bar-Haim, Cohan,
  and Shmueli-Scheuer]{yehudai2026survey}
Asaf Yehudai, Lilach Eden, Alan Li, Guy Uziel, Yilun Zhao, Roy Bar-Haim, Arman
  Cohan, and Michal Shmueli-Scheuer.
\newblock A survey on evaluation of {LLM}-based agents.
\newblock In \emph{Findings of the Association for Computational Linguistics:
  {ACL} 2026}, pp.\  26690--26714, San Diego, California, United States, 2026.
  Association for Computational Linguistics.
\newblock \doi{10.18653/v1/2026.findings-acl.1330}.
\newblock URL \url{https://aclanthology.org/2026.findings-acl.1330/}.

\bibitem[Yu et~al.(2023)Yu, Lin, Yu, and Xing]{yu2023gptfuzzer}
Jiahao Yu, Xingwei Lin, Zheng Yu, and Xinyu Xing.
\newblock {GPTFUZZER}: Red teaming large language models with auto-generated
  jailbreak prompts.
\newblock arXiv preprint arXiv:2309.10253, 2023.
\newblock URL \url{https://arxiv.org/abs/2309.10253}.

\bibitem[Zhang et~al.(2025{\natexlab{a}})Zhang, Huang, Mei, Yao, Wang, Zhan,
  Wang, and Zhang]{zhang2025asb}
Hanrong Zhang, Jingyuan Huang, Kai Mei, Yifei Yao, Zhenting Wang, Chenlu Zhan,
  Hongwei Wang, and Yongfeng Zhang.
\newblock Agent security bench ({ASB}): Formalizing and benchmarking attacks
  and defenses in {LLM}-based agents.
\newblock In \emph{International Conference on Learning Representations}, pp.\
  35331--35366, 2025{\natexlab{a}}.
\newblock URL
  \url{https://proceedings.iclr.cc/paper_files/paper/2025/hash/5750f91d8fb9d5c02bd8ad2c3b44456b-Abstract-Conference.html}.

\bibitem[Zhang et~al.(2026)Zhang, Wang, Chen, He, Feng, and
  Yang]{zhang2026memoattack}
Junke Zhang, Jianwei Wang, Sishuo Chen, Yizhang He, Qingshuai Feng, and Zhengyi
  Yang.
\newblock Evolving skill-structured attack memory enhances {LLM} jailbreaking.
\newblock \emph{arXiv preprint arXiv:2605.29237}, 2026.
\newblock \doi{10.48550/arXiv.2605.29237}.
\newblock URL \url{https://arxiv.org/abs/2605.29237}.
\newblock Under review.

\bibitem[Zhang et~al.(2025{\natexlab{b}})Zhang, He, Cai, Ye, Zhao, Feng, and
  Wang]{zhang2025genesis}
Zheng Zhang, Jiarui He, Yuchen Cai, Deheng Ye, Peilin Zhao, Ruili Feng, and Hao
  Wang.
\newblock {Genesis}: Evolving attack strategies for {LLM} web agent
  red-teaming.
\newblock \emph{arXiv preprint arXiv:2510.18314}, 2025{\natexlab{b}}.
\newblock \doi{10.48550/arXiv.2510.18314}.
\newblock URL \url{https://arxiv.org/abs/2510.18314}.
\newblock Accepted by IEEE ICME 2026.

\bibitem[Zhou et~al.(2025)Zhou, Wu, Pinto, Chen, Zeng, Yang, Yang, Koyejo, Zou,
  and Li]{zhou2025autoredteamer}
Andy Zhou, Kevin Wu, Francesco Pinto, Zhaorun Chen, Yi~Zeng, Yu~Yang, Shuang
  Yang, Sanmi Koyejo, James Zou, and Bo~Li.
\newblock {AutoRedTeamer}: Autonomous red teaming with lifelong attack
  integration.
\newblock In \emph{Advances in Neural Information Processing Systems},
  volume~38, pp.\  169852--169895. Curran Associates, Inc., 2025.
\newblock URL
  \url{https://proceedings.neurips.cc/paper_files/paper/2025/hash/f810a445357d94070669d970a95fc5d8-Abstract-Conference.html}.

\bibitem[Zou et~al.(2023)Zou, Wang, Carlini, Nasr, Kolter, and
  Fredrikson]{zou2023gcg}
Andy Zou, Zifan Wang, Nicholas Carlini, Milad Nasr, J.~Zico Kolter, and Matt
  Fredrikson.
\newblock Universal and transferable adversarial attacks on aligned language
  models.
\newblock arXiv preprint arXiv:2307.15043, 2023.
\newblock URL \url{https://arxiv.org/abs/2307.15043}.

\end{thebibliography}
\bibliographystyle{template-iclr2027/iclr2027_conference}


\end{document}